\documentclass[twocolumn,aps,prl,superscriptaddress,nofootinbib]{revtex4-2}
\usepackage{graphicx}
\usepackage{subcaption}
\usepackage{color}
\graphicspath{{figures/}{fig/}}
\usepackage{amsmath}
\usepackage{amssymb}
\usepackage{bm}
\usepackage{slashed}
\usepackage{epsfig}
\usepackage{amsfonts}
\usepackage{epstopdf}
\usepackage{hyperref}
\usepackage{bbm}
\usepackage{textcomp}
\usepackage{color}
\newcommand{\sect}[1]{{\it \textbf{#1.} --- }}

\begin{document}
\title{Heavy-quark pair production at lepton colliders at NNNLO in QCD}

\author{Xiang Chen}
\email{xchenphy@pku.edu.cn}
\affiliation{School of Physics, Peking University, Beijing 100871, China}

\author{Xin Guan}
\email{guanxin0507@pku.edu.cn}
\affiliation{School of Physics, Peking University, Beijing 100871, China}
\affiliation{Center for High Energy Physics, Peking University, Beijing 100871, China}

\author{Chuan-Qi He}
\email{legend\_he@pku.edu.cn}
\affiliation{Key Laboratory of Atomic and Subatomic Structure and Quantum Control (MOE), Guangdong Basic Research Center of Excellence for Structure and Fundamental Interactions of Matter, Institute of Quantum Matter, South China Normal University, Guangzhou 510006, China}
\affiliation{School of Physics, Peking University, Beijing 100871, China}

\author{Xiao Liu}
\email{xiao.liu@physics.ox.ac.uk}
\affiliation{Rudolf Peierls Centre for Theoretical Physics, Clarendon Laboratory, Parks Road, Oxford OX1 3PU, United Kingdom}

\author{Yan-Qing Ma}
\email{yqma@pku.edu.cn}
\affiliation{School of Physics, Peking University, Beijing 100871, China}
\affiliation{Center for High Energy Physics, Peking University, Beijing 100871, China}

\date{\today}

\begin{abstract}

We compute the total cross-section and invariant mass distribution for heavy-quark pair production in $e^+e^-$ annihilation at the next-to-next-to-next-to-leading order in QCD. The obtained results are expressed as piecewise functions defined by several deeply expanded power series, facilitating a rapid numerical evaluation. Utilizing top-pair production at a collision energy of 500 GeV as a benchmark, we observe a correction of approximately $0.1\%$ for the total cross-section and around $10\%$ for the majority of the invariant mass distribution range. These results play a crucial role in significantly reducing theoretical uncertainty: the scale dependence has been diminished to $0.06\%$ for the total cross-section and to $5\%$ for the invariant mass distribution. This reduction of uncertainty meets the stringent requirements of future lepton colliders.
\end{abstract}

\maketitle
\allowdisplaybreaks

%%%%%%%%%%%%%%%%%%%%%%%%%%%%%%%%%%%%%%%%%%%%%%%%%
\sect{Introduction}
As the heaviest particle in the Standard Model of particle physics, the top quark plays a crucial role in both precision electroweak physics and physics beyond the Standard Model~\cite{Schwienhorst:2022yqu}. Currently, the top quark is mainly studied by hadron colliders, which, for example, determine top-quark mass with uncertainties about 500-600 MeV~\cite{ATLAS:2018fwq, CMS:2015lbj}.

Study of the top-quark pair production at future lepton colliders is important to further pin down properties of the top quark~\cite{Baer:2013cma, Bambade:2019fyw, FCC:2018evy, CEPCStudyGroup:2018ghi,Vos:2016til,CLICdp:2018esa,ILDConceptGroup:2020sfq}, due to the substantially cleaner hadronic environment aspects related to hadronic initial state radiation.
For example, at the International Linear Collider~\cite{ILDConceptGroup:2020sfq}, uncertainties of top-quark mass can be reduced to 50 MeV by measuring  top-quark pair production near threshold. While by measuring the
cross-section and forward-backward asymmetry for the top-quark pair production at 500 GeV,  form factors
for the top quark couplings to the photon and the Z boson can be determined
with relative uncertainties smaller than 0.3\%.
To exploit the full potential of future colliders,  it is imperative to undertake advanced computations for the cross-section and differential distributions of top-pair production at higher orders within the framework of perturbation theory.

\begin{figure}[htb]
	\centering
    \begin{minipage}[b]{.35\linewidth}
        \centering
        \includegraphics[width=1.0\linewidth]{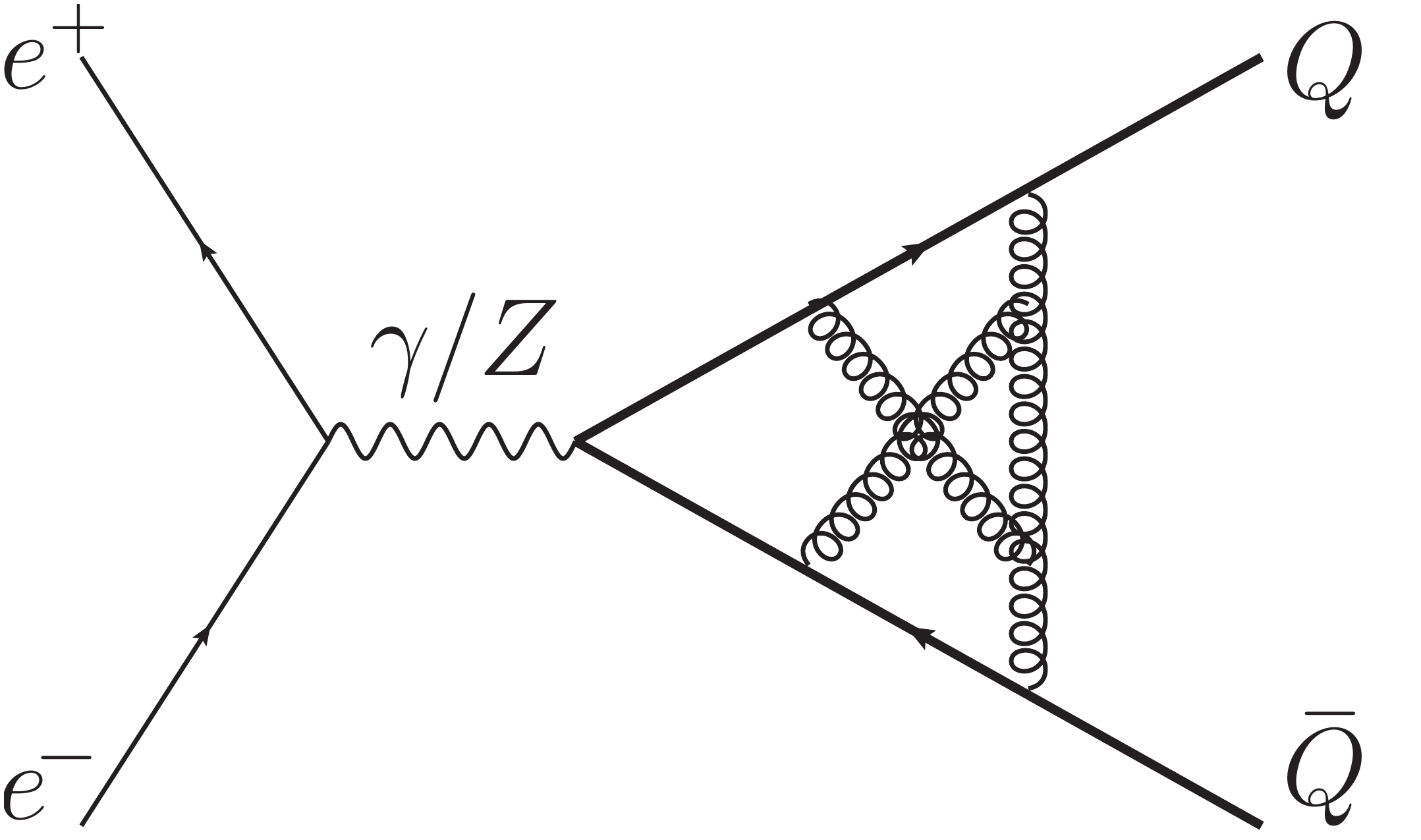}
        \subcaption{}
    \end{minipage}
    \begin{minipage}[b]{.35\linewidth}
        \centering
        \includegraphics[width=1.0\linewidth]{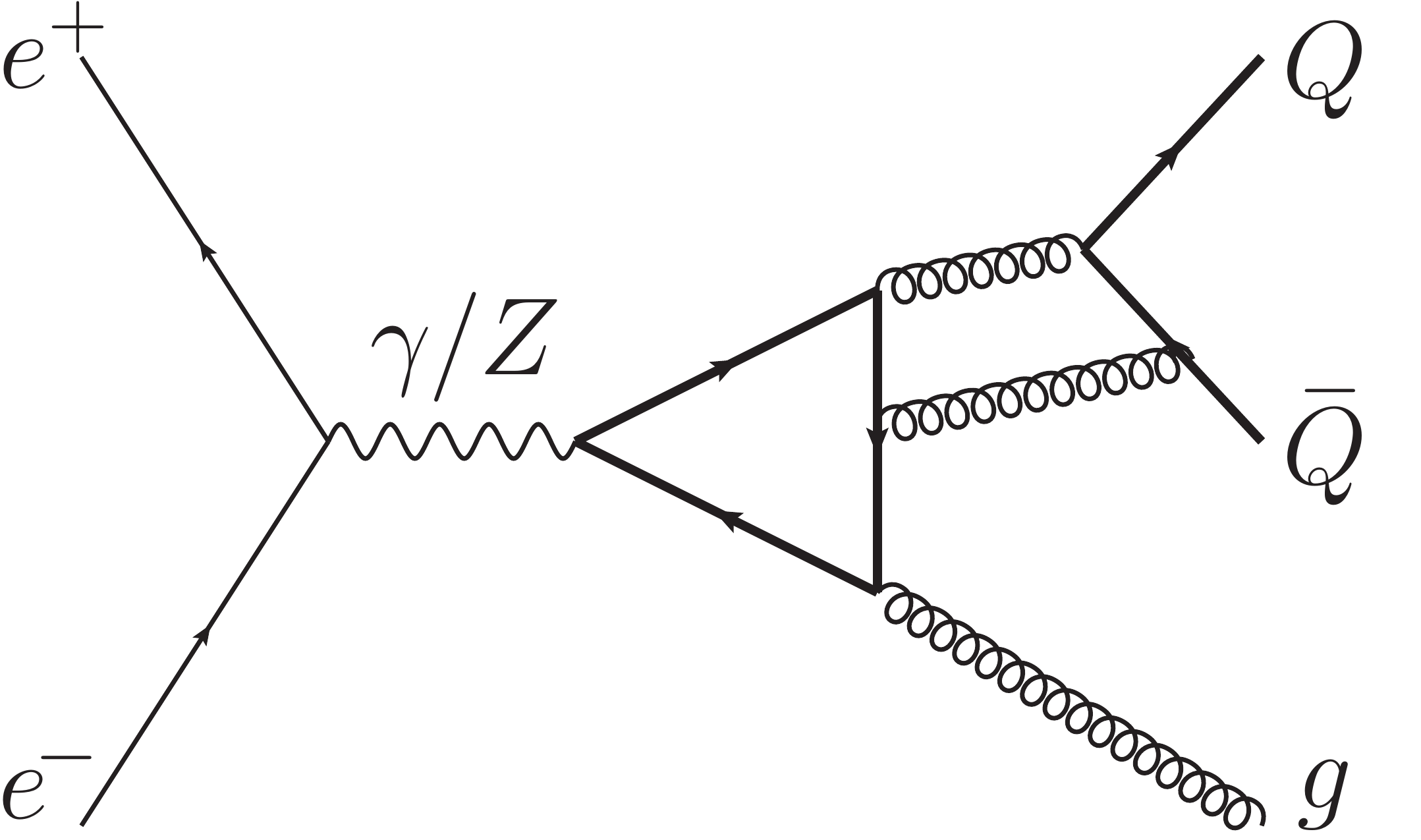}
        \subcaption{}
    \end{minipage}
    \begin{minipage}[b]{.35\linewidth}
        \centering
        \includegraphics[width=1.0\linewidth]{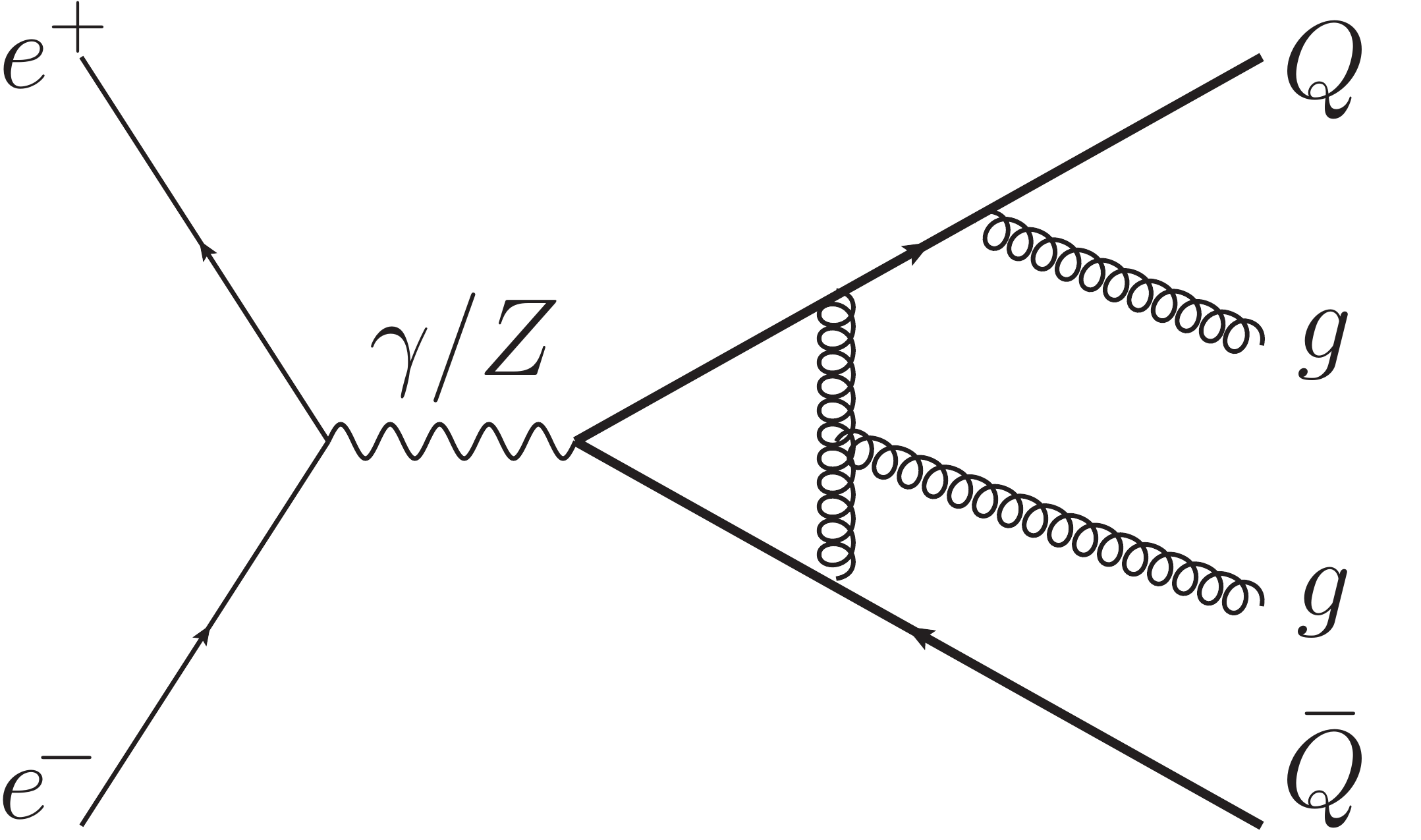}
        \subcaption{}
    \end{minipage}
    \begin{minipage}[b]{.35\linewidth}
        \centering
        \includegraphics[width=1.0\linewidth]{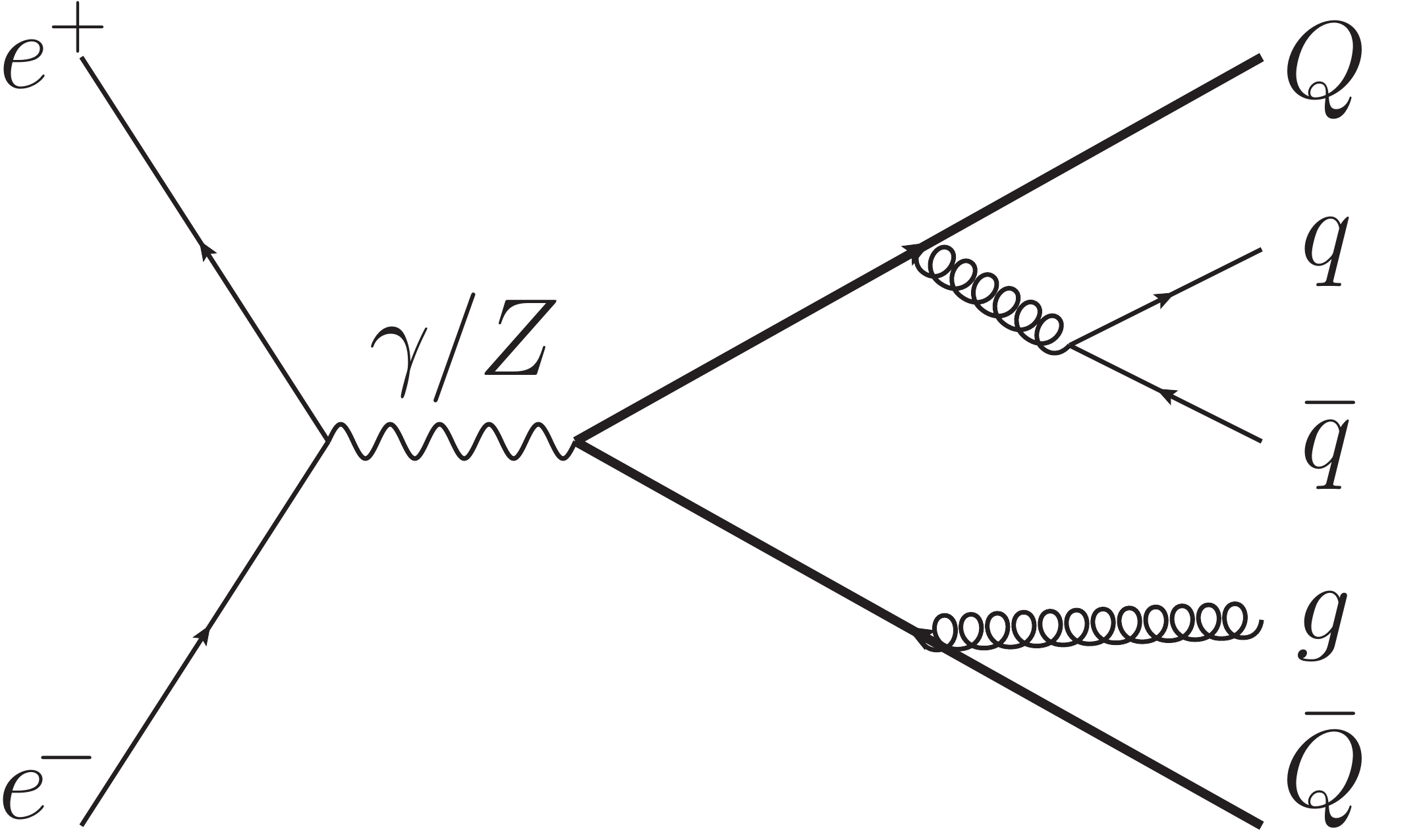}
        \subcaption{}
    \end{minipage}
	\caption{\label{fig:FeynmanDiagram}
			Representative Feynman diagrams at NNNLO.
}
\end{figure}

Full next-to-leading order (NLO) QCD correction was first computed in Ref.~\cite{Jersak:1981sp} and NLO electroweak  effects together with NLO QCD correction was provided in Ref.~\cite{Beenakker:1991ca}. Next-to-next-to-leading order (NNLO) QCD correction to the total cross-section was first obtained in Ref.~\cite{Chetyrkin:1996cf} using Pad$\acute{\text{e}}$ approximation based on the results of threshold expansions, high-energy expansions and %small $s$
low-energy expansions. Direct calculation of the cross-section and differential distributions at NNLO QCD with full top-mass dependence was provided in Refs.~\cite{Gao:2014nva,Gao:2014eea,Chen:2016zbz,Capatti:2022tit}. At the next-to-next-to-next-to-leading order (NNNLO) level, partial total cross-section was obtained utilizing Pad$\acute{\text{e}}$ approximation~\cite{Hoang:2008qy, Kiyo:2009gb}, specifically excluding the singlet contribution like diagrams such as Fig.~\ref{fig:FeynmanDiagram} (b). For top-pair production cross-section near threshold, NNNLO QCD correction has been achieved in Refs.~\cite{Beneke:2015kwa, Beneke:2016kkb}. In Refs.~\cite{Fael:2022rgm,Fael:2022miw}, massive form factors at three loops was obtained, which are building blocks for complete NNNLO QCD corrections. Despite significant theoretical advancements, a comprehensive NNNLO QCD calculation for the total cross-section is notably absent. Moreover, tackling the complexities of differential observables remains a substantial challenge, with no single study addressing this in the existing literature.

In this Letter, we offer a comprehensive computation of the NNNLO QCD corrections to the total cross-section in the process $e^+e^-\to\gamma^*/Z^*\to t\bar{t}$. This encompasses scenarios involving an off-shell photon ($\gamma^*$) and/or an off-shell $Z$ boson ($Z^*$), and the validity extends across the entire physical region. Additionally, we introduce a first computation of the differential distribution up to NNNLO in QCD, exemplified by the invariant mass distribution of the top-quark pair. Our computations are achieved in a semi-analytical way taking advantage of newly proposed methods and released tools. All integrals including phase-space integrals and loop integrals are expressed as piecewise functions represented by several deeply expanded power series in the physical region. As a result, the final total cross-section and invariant mass distribution can be used precisely and efficiently.  Employing identical methodologies, we further derive NNNLO QCD corrections for the production cross-sections of $c\bar{c}$ and $b\bar{b}$.

%%%%%%%%%%%%%%%%%%%%%%%%%%%%%%%%%%%%%%%%%%%%%%%%%
\sect{Calculation}
We take the total cross-section as an example to illustrate our computation methods. We generate the Feynman amplitudes with {\tt qgraf}~\cite{Nogueira:1991ex}, with some sample Feynman diagrams shown in Fig.~\ref{fig:FeynmanDiagram}. Note that the four-top final state contributions, which can be measured separately, are not included in our calculation. We use our in-house {\tt Mathematica} package to deal with Lorentz and color algebras, and express all squared amplitudes as linear combinations of scalar integrals, which have been mapped to predefined integral families respectively. Phase-space integrals are treated in the same manner as loop integrals in our calculation taking advantage of the reverse unitarity~\cite{Anastasiou:2002yz,Anastasiou:2002qz,Anastasiou:2003yy}. The coefficients of integrals are some simple rational functions of the squared center-of-mass energy $s$, squared heavy quark mass $m^2$, dimensional regulator $\epsilon=(4-D)/2$, number of light fermions $n_l$ and $N_c$ of the $SU(N_c)$ gauge group.

Integration-by-parts (IBP) reduction~\cite{Chetyrkin:1981qh} is used to express integrals in each family as linear combinations of a minimal set of so-called master integrals. We  realize this by a two-step strategy. At the first step, we use {\tt LiteRed}~\cite{Lee:2013mka} and {\tt FiniteFlow}~\cite{Peraro:2019svx} to generate and solve the system of IBP identities based on Laporta's algorithm~\cite{Laporta:2000dsw} over finite field. A small number (typically 50) of numerical samplings are sufficient for us to construct the block-triangular relations for target integrals proposed in Refs. ~\cite{Liu:2018dmc, Guan:2019bcx}.
At the second step, we use these much more efficient block-triangular relations to obtain large amount (typically 1000) of samplings to eventually reconstruct the reduction coefficients. In this work, the two-step strategy typically reduces the computational time by one to two orders, depending on families, compared with the reduction without using block-triangular relations.  We note that this is the first time the block-triangular relations are used to tackle a physical problem.

The master integrals are computed using differential equations~\cite{Kotikov:1990kg} based on power series expansion~\cite{Caffo:2008aw, Czakon:2008zk}. The differential equations of master integrals with respect to $x=4m^2/s$ are constructed using the aforementioned IBP reduction. To fix the boundary conditions, say at an arbitrary regular point $x=4/23$, we utilize the auxiliary mass flow method~\cite{Liu:2017jxz,Liu:2020kpc, Liu:2021wks, Liu:2022mfb} implemented in {\tt AMFlow}~\cite{Liu:2022chg}.
With these in hand, we are able to construct a piecewise function for each master integral using the differential equation solver in {\tt AMFlow}, as 5 deeply expanded power series at the following points $x \to\left\{ 0, \frac{1}{4}, \frac{1}{2},\frac{3}{4}, 1 \right\}$,
where $0$, $1/4$ and $1$ are singular points and $1/2$ and $3/4$ are regular points. The expansion at each point, say $x_0$, offers a rapidly converging estimation, particularly within the range $x\in(x_0-1/8,x_0+1/8)$. Consequently, the entire physical region $x\in(0,1)$ is comprehensively covered.

It is widely recognized that within dimensional regularization, the simultaneous satisfaction of the anticommutation relation $\{\gamma^\mu, \gamma_5\}=0$ and the cyclicity of the Dirac trace poses a challenge. In practical applications, the maintenance of the anticommutation relation is not only favored for computational simplification but is also valued for its potential to safeguard chiral symmetry and, consequently, gauge invariance~\cite{Jegerlehner:2000dz}. This approach necessitates a specific prescription for reading a fermion loop, which we choose the KKS scheme~\cite{Kreimer:1989ke, Korner:1991sx, Kreimer:1993bh}. In the KKS scheme, the ultimate result is defined as the average of all conceivable reading points, originating from both the head and tail of an axial-vector three-point fermion-gauge subgraph that encompasses the maximal one-particle-irreducible non-singlet-type loop correction~\cite{Kreimer:1993bh, Chen:2023lus}. In this context, the terms ``singlet" and ``non-singlet" are defined within the framework of Feynman diagrams, indicating whether the external current directly couples to a fermion loop within the relevant part of the Feynman diagram. This distinction is exemplified in Fig.~\ref{fig:FeynmanDiagram} (b) and \ref{fig:FeynmanDiagram}(a, c, d), respectively.
For our calculation within the Standard Model, which is anomaly-free, no additional finite renormalization is required to restore symmetries when using the KKS scheme, at least to the order considered here~\cite{Kreimer:1993bh, Chen:2023lus}.

As we are working in the bare perturbation theory, we need to replace bare quantities by renormalized ones before physical results can be obtained. QCD coupling is renormalized in the $\overline{\text{MS}}$ scheme, and  heavy quark mass and fields are renormalized in the on-shell scheme. Both coupling renormalization and field renormalization can be realized by simply multiplying the bare results by proper powers of $Z_{\alpha_s}^{\overline{\text{MS}}}$ and $Z_2^{\text{OS}}$. Heavy quark mass renormalization is more complicated because it is not multiplicatively renormalizable.
We replace bare mass by renormalized mass via $m^b=m^{\text{OS}}+\delta m$, and keep on-shell condition of external heavy quark momentum $p^2=(m^{\text{OS}})^2$ untouched.
In the spirit of perturbation theory, $\delta m$ is a small quantity and can be expanded to any desired order.
The expansion can be readily implemented either at the integrand level or, alternatively, at the integral level by formulating and solving differential equations with respect to $\delta m$. After the expansion of $\delta m$, we can substitute  renormalization constants, which can be found in Refs.~\cite{Melnikov:2000zc,Czakon:2007wk,Barnreuther:2013qvf}, to obtain the final renormalized cross-section. We note that performing renormalization in this way makes our calculation very systematic, and at the same time introduces negligible efforts against the calculation of bare cross-section at the highest order in $\alpha_s$.

Another key technique in our calculation is that the master integrals are calculated with numerical values of $\epsilon$ and the $\epsilon$ dependence is only reconstructed at the cross-section level, as proposed in Refs.~\cite{Liu:2022mfb,Liu:2022chg}. The advantage of this technique is that we do not need to manipulate Laurent expansions of $\epsilon$ during the intermediate stages of calculations, which significantly reduces the computational time.

Finally, we discuss how to compute differential cross-sections based on the previously outlined methods. Using the invariant mass distribution of the heavy-quark pair as an example, the differential cross-section can be derived by inserting a delta function $\delta \left( (p_Q + p_{\bar{Q}})^2 - M_{Q\bar{Q}}^2 \right)$ into the final state phase space integration.
This integration can be approached similarly to the total cross-section calculation, noting that the delta function can be expressed as cut propagators using reverse unitarity~\cite{Anastasiou:2002yz,Anastasiou:2002qz,Anastasiou:2003yy}. We highlight that the boundary conditions for differential equations of master integrals in differential cross-section computation can be determined by aligning them with integrals from the total cross-section. Specifically, we solve the differential equation and formulate master integrals as piecewise functions of $M_{Q \bar{Q}}$, represented by 13 deeply expanded power series with undetermined coefficients. By integrating over $M_{Q \bar{Q}}$, we collapse the cut propagator and arrive at integrals that belong to the integral families of the total cross-section, which are already known. Consequently, we can ascertain both the unknown coefficients and the piecewise functions.

Thanks to all these strategies mentioned above, the computational resources utilized in this work amount to less than $10^5$ CPU core hours in total.

%%%%%%%%%%%%%%%%%%%%%%%%%%%%%%%%%%%%%%%%%%%%%%%%%
\sect{Results}
By combining everything together, we end up with final results at the NNNLO level, which are free of ultraviolet and infrared divergences as expected from the Kinoshita-Lee-Nauenberg theorem \cite{Kinoshita:1962ur,Lee:1964is}.
Our result of the NNNLO total cross-section is expressed as a piecewise function of $x=4m^2/s$ represented by 5 power series. Expanding these series to 40 orders enables us to achieve at least 10 correct digits in the physical region, with a relative error of approximately $2^{-40}\sim 10^{-12}$. The expressions can be found in a computer-readable ancillary file attached to this Letter.

To provide numerical result for top-quark pair production, we choose top-quark mass as $m=172.69$ GeV \cite{ParticleDataGroup:2022pth} and set all other quarks as massless. Electromagnetic coupling is chosen as a fixed value
$\alpha=1/132.2$, and strong coupling $\alpha_s(\mu)$ is running as a function of renormalization scale $\mu$, which is computed using the {\tt RunDec} package~\cite{Chetyrkin:2000yt, Herren:2017osy} with input value $\alpha_s^{n_f=5}(m_Z)=0.1181$. Other electroweak parameters are chosen from Ref.
\cite{ParticleDataGroup:2022pth}.

\begin{figure}[htb]
		\includegraphics[width=0.85\linewidth]{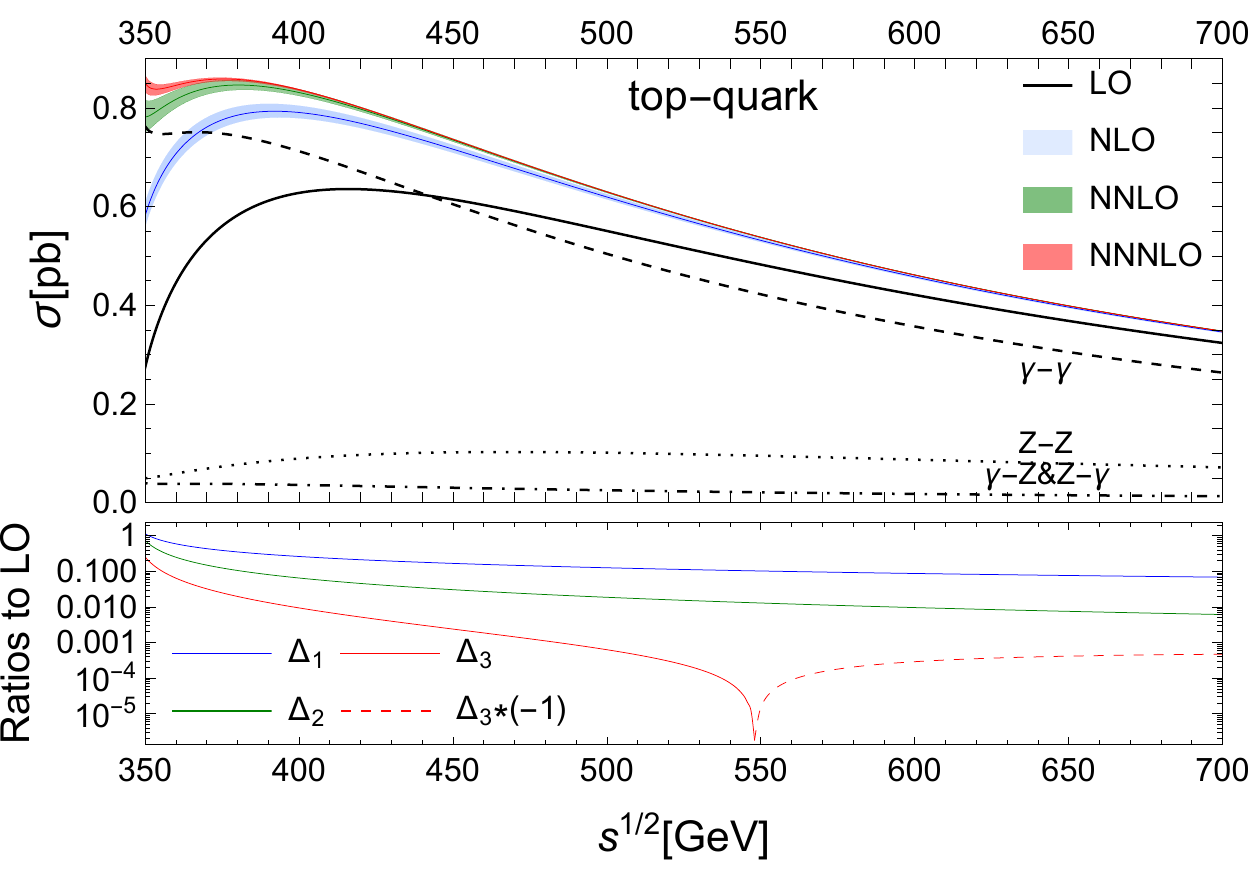}
		\caption{\label{fig:TCSTop}
Total cross-section for $t\bar{t}$ production.
Refer to the text for details.
		}
\end{figure}

Fig.~\ref{fig:TCSTop} shows our result for the total cross-section of $t\bar t$ production, where LO, NLO, NNLO and NNNLO are shown in black, blue, green and red respectively, and the dashed, dotted, and dot-dashed lines correspond to $\gamma-\gamma$, $Z-Z$, $\gamma-Z$ and $Z-\gamma$ contributions at NNNLO, respectively. In the upper panel, the middle lines of each band correspond to the choice of $\mu=\sqrt{s}$ for the renormalization scale, and the upper and lower lines correspond to the scale variations between $\mu=\sqrt{s}/2$ and $\mu=2\sqrt{s}$. It can be found that the NNNLO correction significantly reduces the scale dependence. However, the NNNLO result near the production threshold, say for $\sqrt{s}<370$ GeV,  still suffers from large uncertainty due to Coulomb interaction. Perturbative calculations become unreliable in this region, necessitating the application of resummation for further improvement~\cite{Hoang:2013uda,Beneke:2015kwa,Beneke:2016kkb}.

The total cross-section can also be expressed in the form
\begin{equation}\label{equ:DeltaRatio}
  \sigma_{\text{NNNLO}}=\sigma_{\text{LO}}(1+\Delta_1+\Delta_2+\Delta_3),
\end{equation}
where the order $\alpha_s$, $\alpha_s^2$ and $\alpha_s^3$ corrections $\Delta_1$, $\Delta_2$ and $\Delta_3$ are displayed in the lower panel of Fig.~\ref{fig:TCSTop} as functions of the center-of-mass energy, respectively. The renormalization scale is set to $\sqrt{s}$. The smallness of $\Delta_3$ confirms a good convergence of perturbative expansion with respect to $\alpha_s$.

\begin{figure}[htb]
	\begin{center}
		\includegraphics[width=0.85\linewidth]{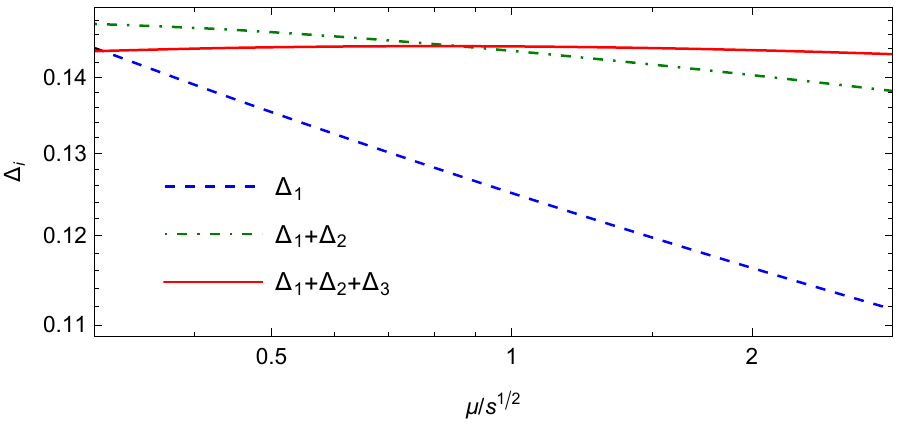}
		\caption{\label{fig:MD500}
			Scale dependence for $t\bar{t}$ production at $\sqrt{s}=500$ GeV.}
	\end{center}
\end{figure}

In Fig.~\ref{fig:MD500}, we further show the reduction of the scale dependence after including $O(\alpha_s^3)$ correction by varying the renormalization scale in a larger range. It is found that, for a collision energy of 500 GeV, the scale dependence has been reduced from $0.60\%$ at NNLO to $0.06\%$ at NNNLO by varying the scale between $\mu=\sqrt{s}/2$ and $\mu=2\sqrt{s}$, which meets the precision requested by, e.g., International Linear Collider~\cite{ILDConceptGroup:2020sfq}.

\begin{figure}[htb]
	\begin{center}
		\includegraphics[width=0.85\linewidth]{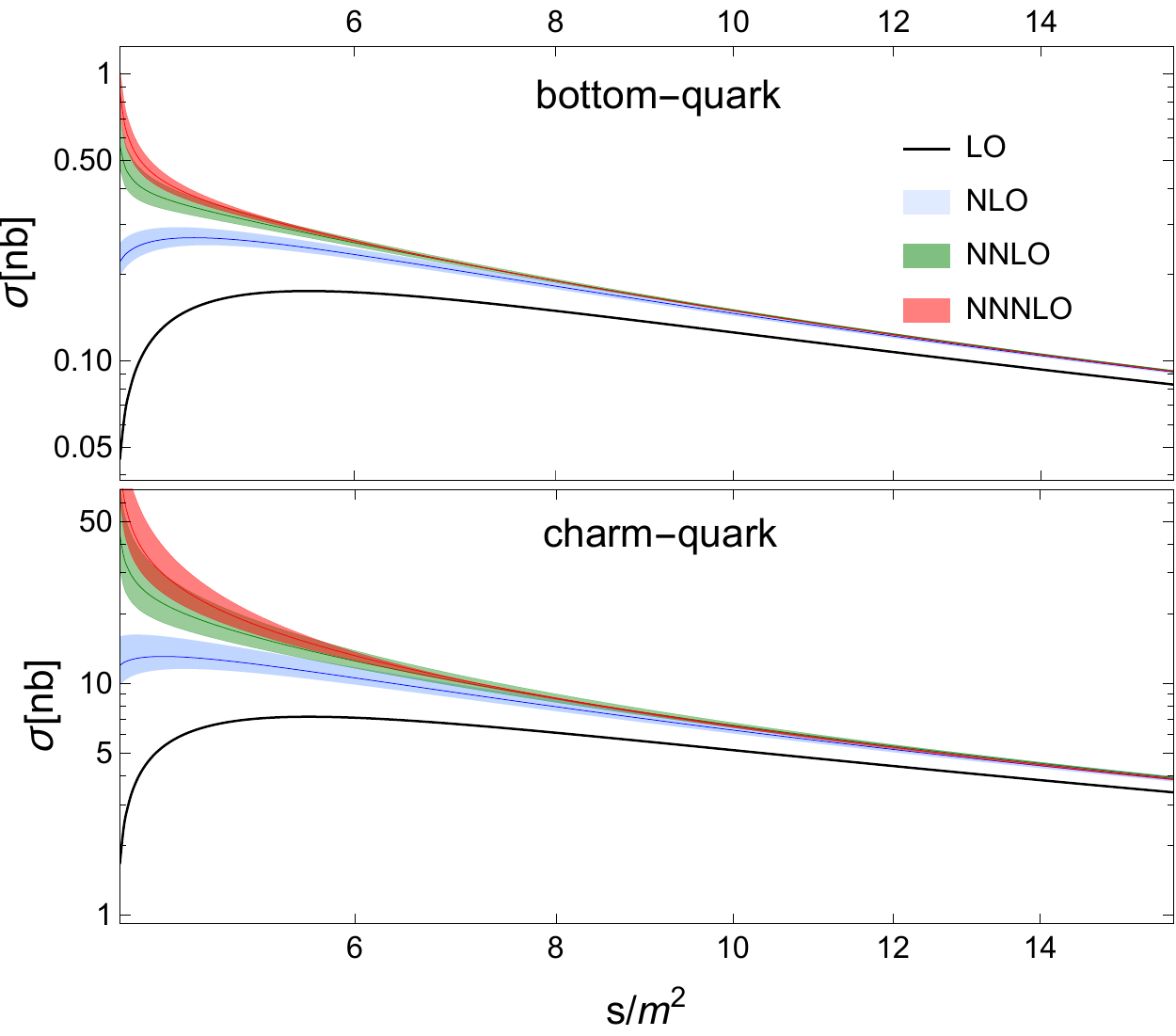}
		\caption{\label{fig:TCSBottomCharm}
			Total cross-section for $b\bar{b}$ and $c\bar{c}$ production.
		}
	\end{center}
\end{figure}

The situation is analogous for $b\bar b$ or $c\bar c$ production cases, as illustrated in Fig.~\ref{fig:TCSBottomCharm}. In these instances, an on-shell heavy-quark mass is chosen as $m_b=4.8$ GeV and $m_c=1.5$ GeV, respectively. Except the threshold region where perturbation is unreliable, theoretical uncertainties are significantly reduced, making them valuable for describing corresponding processes, like the so-called $R$ ratio. It is important to note that, in our calculation, quarks with masses heavier than the considered heavy quark are excluded, while those with masses lighter than the considered quark are treated as massless. Additionally, our consideration is limited to diagrams mediated by an off-shell photon ($\gamma^*$) since the mass of the $Z$ boson significantly exceeds the energy scale of interest here.

\begin{figure}[htb]
	\begin{center}
		\includegraphics[width=0.85\linewidth]{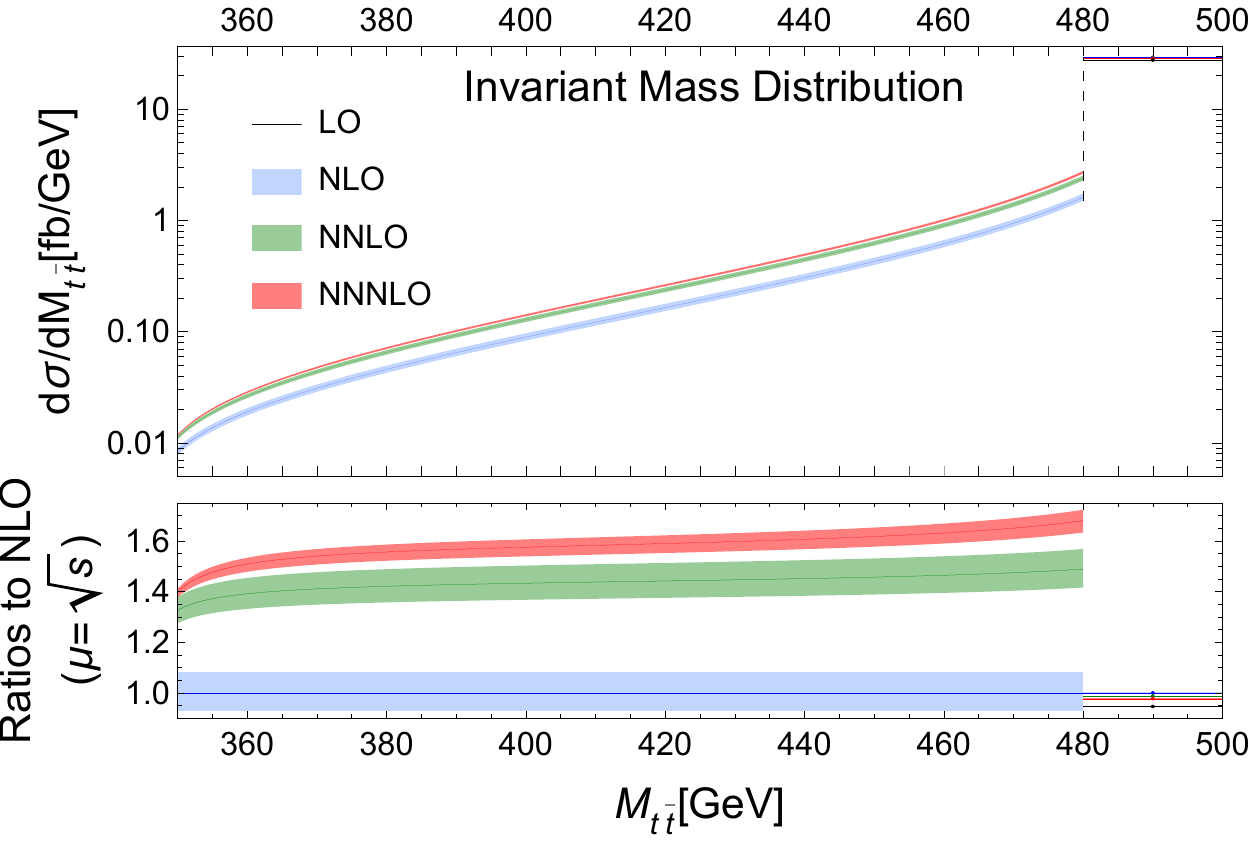}
		\caption{\label{fig:MttDist500}
			The $M_{t\bar{t}}$ distribution at $\sqrt{s}=500$ GeV.}
	\end{center}
\end{figure}

As a first instance of NNNLO differential cross-sections, Fig.~\ref{fig:MttDist500} illustrates the invariant mass distribution of the top-quark pair for a collision energy of 500 GeV. The distribution increases as $M_{t\bar{t}}$ becomes larger and becomes singular when $M_{t\bar{t}}$ approaches its maximum value of $\sqrt{s}$, where QCD radiations must be soft. In this region, the fixed-order prediction breaks down, prompting us to take an average over a 20 GeV bin.
Although fixed-order perturbation theory also breaks down when $M_{t\bar{t}}\sim 2m_t$, this region is negligible due to the significant suppression by phase space considerations.
On the whole, the pure $\mathcal{O}(\alpha_s^3)$ correction introduces a significant modification to the lower order result,  falling within the range of approximately 9\% to 13\% for $M_{t\bar{t}} \in [370,480]$ GeV and about $-1\%$ for the bin at the end point. The scale uncertainties have been mitigated to approximately $5\%$ across the majority of the distribution range. This represents a considerable impact for phenomenological study.

%%%%%%%%%%%%%%%%%%%%%%%%%%%%%%%%%%%%%%%%%%%%%%%%%
\sect{Verification}
Besides free of ultraviolet and infrared divergences, our results have been checked in several other aspects, which we describe in the following.

Our result for $t\bar{t}$ total cross-section and invariant mass distribution at NNLO agrees with that obtained in Refs.~\cite{Gao:2014nva,Chen:2016zbz}. Specifically, we have replicated Fig. 3, Fig. 7, and the left plot of Fig. 10 in Ref.~\cite{Gao:2014nva}, and reproduced Fig. 1, Fig. 2, and the right plot of Fig. 5 in Ref.~\cite{Chen:2016zbz}.  Our three-loop contribution agrees with Refs.~\cite{Fael:2022rgm,Fael:2022miw} to 7 digits provided there. Our non-singlet results at NNNLO agree with~\cite{Hoang:2008qy,Kiyo:2009gb} within their uncertainty.\footnote{While singlet contributions are suppressed near the production threshold, as they exchange two fewer Coulomb gluons compared to non-singlet contributions, they become crucial with increasing energy. Numerically, NNNLO singlet contributions become as large as non-singlet contributions when $\sqrt{s}/(2m) > 5$.} All master integrals involved in total cross-section,  obtained by solving differential equations with respect to $x$ with boundary conditions given at $x=4/23$, have been checked by {\tt AMFlow} at another phase space point $x=15/16$ with at least 10 digits precision. By integrating the invariant mass differential distribution at NNNLO, we found perfect agreement with total cross-section.

For the vector current contributions, we also calculate the forward scattering amplitude of a virtual photon at 4-loop level without cut, and then relate its imaginary part to the desired cross-section by virtue of optical theorem. The results agree perfectly with that obtained using default methods.
Note that using optical theorem could induce undesired final states, including states with no heavy quarks and states with four heavy quarks, which are subtracted by calculating virtual corrections and real radiations using the default methods.

To validate the appropriateness of the chosen $\gamma_5$ scheme, we performed a cross-check by computing the singlet part of the axial-axial contribution at NNNLO using the Larin scheme~\cite{Larin:1991tj,Larin:1993tq}. Our focus was specifically on the axial-axial contribution, as the vector-vector contribution does not involve $\gamma_5$, and the vector-axial contribution is zero.
Moreover, for the non-singlet portion of the axial-axial contribution, the use of the anticommuting $\gamma_5$ scheme is unambiguous, because there will always be an even number of $\gamma_5$ in the Dirac trace. Consequently, our consideration is confined to the singlet part of the axial-axial contribution.
Our findings indicate that the results obtained using the KKS scheme align with those obtained using the Larin scheme, following a finite renormalization in the latter scheme~\cite{Larin:1991tj,Larin:1993tq,Ahmed:2021spj}.

%%%%%%%%%%%%%%%%%%%%%%%%%%%%%%%%%%%%%%%%%%%%%%%%%
\sect{Summary}
In this Letter, we present the complete NNNLO QCD correction to the cross-section of heavy-quark pair production at lepton colliders. Additionally, we provide the first differential cross-section, specifically the invariant mass distribution, at the same order. Using top-quark pair production at a collision energy of 500 GeV as a benchmark, we observe a correction of approximately 0.1\% for the total cross-section, while the correction for the invariant mass distribution ranges around 10\% for most values away from the end point. The NNNLO QCD correction largely reduces the scale dependence. These results hold significant value for the precision testing of heavy-quark pair production.

The derivation of our results involves the integration of various powerful techniques, like reverse unitarity, finite field, block-triangular relations, and numerical sampling of $\epsilon$. Our strategic approach is not only tailored to the specific process studied here but is also versatile enough to be applied to numerous other processes, including differential observables. This versatility holds promise for advancing perturbative calculations to unprecedented orders in various contexts.

%%%%%%%%%%%%%%%%%%%%%%%%%%%%%%%%%%%%%%%%%%%%%%%%%
\begin{acknowledgments}
	\sect{Acknowledgments}
	We thank L. Chen and R.J. Huang for many useful communications and discussions.
	The work was supported in part by the National Natural Science Foundation of
	China (Grants No. 12325503, No. 11975029, No. 11875071, ), the National Key Research and Development Program of China under
	Contracts No. 2020YFA0406400, and the High-performance Computing Platform of Peking University. The research of X.L. was also supported by the ERC Starting Grant 804394 \textsc{HipQCD} and by the UK Science and Technology Facilities Council (STFC) under Grant No. ST/T000864/1.
    {\tt JaxoDraw}~\cite{BINOSI200476} was used to generate Feynman diagrams.
\end{acknowledgments}

%%%%%%%%%%%%%%%%%%%%%%%%%%%%%%%%%%%%%%%%%%%%%%%%%
%%%%%%%%%%%%%%%%%%%%%%%%%%%%%%%%%%%%%%%%%%%%%%%%%
% references
%\bibliographystyle{utphysMa}
%\bibliography{bibTex1.5}

\begin{thebibliography}{10}

\bibitem{Schwienhorst:2022yqu}
K.~Agashe {\em et al.}, {\it {Report of the Topical Group on Top quark physics
  and heavy flavor production for Snowmass 2021}},
  [\href{http://arxiv.org/abs/2209.11267}{{\ttfamily arXiv:2209.11267}}]
  [\href{http://inspirehep.net/search?p=find+Schwienhorst:2022yqu}{{\ttfamily
  InSPIRE}}].

\bibitem{ATLAS:2018fwq}
{\bfseries ATLAS} , M.~Aaboud {\em et al.}, {\it {Measurement of the top quark
  mass in the $t\bar{t}\rightarrow $ lepton+jets channel from $\sqrt{s}=8$ TeV
  ATLAS data and combination with previous results}},
  \href{http://dx.doi.org/10.1140/epjc/s10052-019-6757-9}{{\em Eur. Phys. J. C}
  {\bfseries 79} (2019) 290} [\href{http://arxiv.org/abs/1810.01772}{{\ttfamily
  arXiv:1810.01772}}]
  [\href{http://inspirehep.net/search?p=find+ATLAS:2018fwq}{{\ttfamily
  InSPIRE}}].

\bibitem{CMS:2015lbj}
{\bfseries CMS} , V.~Khachatryan {\em et al.}, {\it {Measurement of the top
  quark mass using proton-proton data at ${\sqrt{(s)}}$ = 7 and 8 TeV}},
  \href{http://dx.doi.org/10.1103/PhysRevD.93.072004}{{\em Phys. Rev. D}
  {\bfseries 93} (2016) 072004}
  [\href{http://arxiv.org/abs/1509.04044}{{\ttfamily arXiv:1509.04044}}]
  [\href{http://inspirehep.net/search?p=find+CMS:2015lbj}{{\ttfamily
  InSPIRE}}].

\bibitem{Baer:2013cma}
{\it {The International Linear Collider Technical Design Report - Volume 2:
  Physics}},  [\href{http://arxiv.org/abs/1306.6352}{{\ttfamily
  arXiv:1306.6352}}]
  [\href{http://inspirehep.net/search?p=find+Baer:2013cma}{{\ttfamily
  InSPIRE}}].

\bibitem{Bambade:2019fyw}
P.~Bambade {\em et al.}, {\it {The International Linear Collider: A Global
  Project}},  [\href{http://arxiv.org/abs/1903.01629}{{\ttfamily
  arXiv:1903.01629}}]
  [\href{http://inspirehep.net/search?p=find+Bambade:2019fyw}{{\ttfamily
  InSPIRE}}].

\bibitem{FCC:2018evy}
{\bfseries FCC} , A.~Abada {\em et al.}, {\it {FCC-ee: The Lepton Collider}:
  {Future Circular Collider Conceptual Design Report Volume 2}},
  \href{http://dx.doi.org/10.1140/epjst/e2019-900045-4}{{\em Eur. Phys. J. ST}
  {\bfseries 228} (2019) 261--623}
  [\href{http://inspirehep.net/search?p=find+FCC:2018evy}{{\ttfamily
  InSPIRE}}].

\bibitem{CEPCStudyGroup:2018ghi}
{\bfseries CEPC Study Group} , M.~Dong {\em et al.}, {\it {CEPC Conceptual
  Design Report: Volume 2 - Physics \& Detector}},
  [\href{http://arxiv.org/abs/1811.10545}{{\ttfamily arXiv:1811.10545}}]
  [\href{http://inspirehep.net/search?p=find+CEPCStudyGroup:2018ghi}{{\ttfamily
  InSPIRE}}].

\bibitem{Vos:2016til}
M.~Vos {\em et al.}, {\it {Top physics at high-energy lepton colliders}},
  [\href{http://arxiv.org/abs/1604.08122}{{\ttfamily arXiv:1604.08122}}]
  [\href{http://inspirehep.net/search?p=find+Vos:2016til}{{\ttfamily
  InSPIRE}}].

\bibitem{CLICdp:2018esa}
{\bfseries CLICdp} , H.~Abramowicz {\em et al.}, {\it {Top-Quark Physics at the
  CLIC Electron-Positron Linear Collider}},
  \href{http://dx.doi.org/10.1007/JHEP11(2019)003}{{\em JHEP} {\bfseries 11}
  (2019) 003} [\href{http://arxiv.org/abs/1807.02441}{{\ttfamily
  arXiv:1807.02441}}]
  [\href{http://inspirehep.net/search?p=find+CLICdp:2018esa}{{\ttfamily
  InSPIRE}}].

\bibitem{ILDConceptGroup:2020sfq}
{\bfseries ILD Concept Group} , H.~Abramowicz {\em et al.}, {\it {International
  Large Detector: Interim Design Report}},
  [\href{http://arxiv.org/abs/2003.01116}{{\ttfamily arXiv:2003.01116}}]
  [\href{http://inspirehep.net/search?p=find+ILDConceptGroup:2020sfq}{{\ttfamily
  InSPIRE}}].

\bibitem{Jersak:1981sp}
J.~Jersak, E.~Laermann, and P.~M. Zerwas, {\it {Electroweak Production of Heavy
  Quarks in e+ e- Annihilation}},
  \href{http://dx.doi.org/10.1103/PhysRevD.25.1218}{{\em Phys. Rev. D}
  {\bfseries 25} (1982) 1218}
  [\href{http://inspirehep.net/search?p=find+Jersak:1981sp}{{\ttfamily
  InSPIRE}}]. [Erratum: Phys.Rev.D 36, 310 (1987)].

\bibitem{Beenakker:1991ca}
W.~Beenakker, S.~C. van~der Marck, and W.~Hollik, {\it {e+ e- annihilation into
  heavy fermion pairs at high-energy colliders}},
  \href{http://dx.doi.org/10.1016/0550-3213(91)90606-X}{{\em Nucl. Phys. B}
  {\bfseries 365} (1991) 24--78}
  [\href{http://inspirehep.net/search?p=find+Beenakker:1991ca}{{\ttfamily
  InSPIRE}}].

\bibitem{Chetyrkin:1996cf}
K.~G. Chetyrkin, J.~H. Kuhn, and M.~Steinhauser, {\it {Three loop polarization
  function and O (alpha-s**2) corrections to the production of heavy quarks}},
  \href{http://dx.doi.org/10.1016/S0550-3213(96)00534-2}{{\em Nucl. Phys. B}
  {\bfseries 482} (1996) 213--240}
  [\href{http://arxiv.org/abs/hep-ph/9606230}{{\ttfamily hep-ph/9606230}}]
  [\href{http://inspirehep.net/search?p=find+Chetyrkin:1996cf}{{\ttfamily
  InSPIRE}}].

\bibitem{Gao:2014nva}
J.~Gao and H.~X. Zhu, {\it {Electroweak prodution of top-quark pairs in e+e-
  annihilation at NNLO in QCD: the vector contributions}},
  \href{http://dx.doi.org/10.1103/PhysRevD.90.114022}{{\em Phys. Rev. D}
  {\bfseries 90} (2014) 114022}
  [\href{http://arxiv.org/abs/1408.5150}{{\ttfamily arXiv:1408.5150}}]
  [\href{http://inspirehep.net/search?p=find+Gao:2014nva}{{\ttfamily
  InSPIRE}}].

\bibitem{Gao:2014eea}
J.~Gao and H.~X. Zhu, {\it {Top Quark Forward-Backward Asymmetry in $e^+e^-$
  Annihilation at Next-to-Next-to-Leading Order in QCD}},
  \href{http://dx.doi.org/10.1103/PhysRevLett.113.262001}{{\em Phys. Rev.
  Lett.} {\bfseries 113} (2014) 262001}
  [\href{http://arxiv.org/abs/1410.3165}{{\ttfamily arXiv:1410.3165}}]
  [\href{http://inspirehep.net/search?p=find+Gao:2014eea}{{\ttfamily
  InSPIRE}}].

\bibitem{Chen:2016zbz}
L.~Chen, O.~Dekkers, D.~Heisler, W.~Bernreuther, and Z.-G. Si, {\it {Top-quark
  pair production at next-to-next-to-leading order QCD in electron positron
  collisions}},  \href{http://dx.doi.org/10.1007/JHEP12(2016)098}{{\em JHEP}
  {\bfseries 12} (2016) 098} [\href{http://arxiv.org/abs/1610.07897}{{\ttfamily
  arXiv:1610.07897}}]
  [\href{http://inspirehep.net/search?p=find+Chen:2016zbz}{{\ttfamily
  InSPIRE}}].

\bibitem{Capatti:2022tit}
Z.~Capatti, V.~Hirschi, and B.~Ruijl, {\it {Local unitarity: cutting raised
  propagators and localising renormalisation}},
  \href{http://dx.doi.org/10.1007/JHEP10(2022)120}{{\em JHEP} {\bfseries 10}
  (2022) 120} [\href{http://arxiv.org/abs/2203.11038}{{\ttfamily
  arXiv:2203.11038}}] [\href{http://inspirehep.net/search?p=find+Capatti:2022tit}{{\ttfamily
  InSPIRE}}].

\bibitem{Hoang:2008qy}
A.~H. Hoang, V.~Mateu, and S.~Mohammad~Zebarjad, {\it {Heavy Quark Vacuum
  Polarization Function at O(alpha**2(s)) O(alpha**3(s))}},
  \href{http://dx.doi.org/10.1016/j.nuclphysb.2008.12.005}{{\em Nucl. Phys. B}
  {\bfseries 813} (2009) 349--369}
  [\href{http://arxiv.org/abs/0807.4173}{{\ttfamily arXiv:0807.4173}}]
  [\href{http://inspirehep.net/search?p=find+Hoang:2008qy}{{\ttfamily
  InSPIRE}}].

\bibitem{Kiyo:2009gb}
Y.~Kiyo, A.~Maier, P.~Maierhofer, and P.~Marquard, {\it {Reconstruction of
  heavy quark current correlators at O(alpha(s)**3)}},
  \href{http://dx.doi.org/10.1016/j.nuclphysb.2009.08.010}{{\em Nucl. Phys. B}
  {\bfseries 823} (2009) 269--287}
  [\href{http://arxiv.org/abs/0907.2120}{{\ttfamily arXiv:0907.2120}}]
  [\href{http://inspirehep.net/search?p=find+Kiyo:2009gb}{{\ttfamily
  InSPIRE}}].

\bibitem{Beneke:2015kwa}
M.~Beneke, Y.~Kiyo, P.~Marquard, A.~Penin, J.~Piclum, and M.~Steinhauser, {\it
  {Next-to-Next-to-Next-to-Leading Order QCD Prediction for the Top Antitop
  $S$-Wave Pair Production Cross Section Near Threshold in $e^+e^-$
  Annihilation}},  \href{http://dx.doi.org/10.1103/PhysRevLett.115.192001}{{\em
  Phys. Rev. Lett.} {\bfseries 115} (2015) 192001}
  [\href{http://arxiv.org/abs/1506.06864}{{\ttfamily arXiv:1506.06864}}]
  [\href{http://inspirehep.net/search?p=find+Beneke:2015kwa}{{\ttfamily
  InSPIRE}}].

\bibitem{Beneke:2016kkb}
M.~Beneke, Y.~Kiyo, A.~Maier, and J.~Piclum, {\it {Near-threshold production of
  heavy quarks with $\tt{QQbar\_threshold}$}},
  \href{http://dx.doi.org/10.1016/j.cpc.2016.07.026}{{\em Comput. Phys.
  Commun.} {\bfseries 209} (2016) 96--115}
  [\href{http://arxiv.org/abs/1605.03010}{{\ttfamily arXiv:1605.03010}}]
  [\href{http://inspirehep.net/search?p=find+Beneke:2016kkb}{{\ttfamily
  InSPIRE}}].

\bibitem{Fael:2022rgm}
M.~Fael, F.~Lange, K.~Sch\"onwald, and M.~Steinhauser, {\it {Massive Vector
  Form Factors to Three Loops}},
  \href{http://dx.doi.org/10.1103/PhysRevLett.128.172003}{{\em Phys. Rev.
  Lett.} {\bfseries 128} (2022) 172003}
  [\href{http://arxiv.org/abs/2202.05276}{{\ttfamily arXiv:2202.05276}}]
  [\href{http://inspirehep.net/search?p=find+Fael:2022rgm}{{\ttfamily
  InSPIRE}}].

\bibitem{Fael:2022miw}
M.~Fael, F.~Lange, K.~Sch\"onwald, and M.~Steinhauser, {\it {Singlet and
  nonsinglet three-loop massive form factors}},
  \href{http://dx.doi.org/10.1103/PhysRevD.106.034029}{{\em Phys. Rev. D}
  {\bfseries 106} (2022) 034029}
  [\href{http://arxiv.org/abs/2207.00027}{{\ttfamily arXiv:2207.00027}}]
  [\href{http://inspirehep.net/search?p=find+Fael:2022miw}{{\ttfamily
  InSPIRE}}].

\bibitem{Nogueira:1991ex}
P.~Nogueira, {\it {Automatic Feynman graph generation}},
  \href{http://dx.doi.org/10.1006/jcph.1993.1074}{{\em J. Comput. Phys.}
  {\bfseries 105} (1993) 279--289}
  [\href{http://inspirehep.net/search?p=find+Nogueira:1991ex}{{\ttfamily
  InSPIRE}}].

\bibitem{Anastasiou:2002yz}
C.~Anastasiou and K.~Melnikov, {\it {Higgs boson production at hadron colliders
  in NNLO QCD}},  \href{http://dx.doi.org/10.1016/S0550-3213(02)00837-4}{{\em
  Nucl. Phys. B} {\bfseries 646} (2002) 220--256}
  [\href{http://arxiv.org/abs/hep-ph/0207004}{{\ttfamily hep-ph/0207004}}]
  [\href{http://inspirehep.net/search?p=find+Anastasiou:2002yz}{{\ttfamily
  InSPIRE}}].

\bibitem{Anastasiou:2002qz}
C.~Anastasiou, L.~J. Dixon, and K.~Melnikov, {\it {NLO Higgs boson rapidity
  distributions at hadron colliders}},
  \href{http://dx.doi.org/10.1016/S0920-5632(03)80168-8}{{\em Nucl. Phys. B
  Proc. Suppl.} {\bfseries 116} (2003) 193--197}
  [\href{http://arxiv.org/abs/hep-ph/0211141}{{\ttfamily hep-ph/0211141}}]
  [\href{http://inspirehep.net/search?p=find+Anastasiou:2002qz}{{\ttfamily
  InSPIRE}}].

\bibitem{Anastasiou:2003yy}
C.~Anastasiou, L.~J. Dixon, K.~Melnikov, and F.~Petriello, {\it {Dilepton
  rapidity distribution in the Drell-Yan process at NNLO in QCD}},
  \href{http://dx.doi.org/10.1103/PhysRevLett.91.182002}{{\em Phys. Rev. Lett.}
  {\bfseries 91} (2003) 182002}
  [\href{http://arxiv.org/abs/hep-ph/0306192}{{\ttfamily hep-ph/0306192}}]
  [\href{http://inspirehep.net/search?p=find+Anastasiou:2003yy}{{\ttfamily
  InSPIRE}}].

\bibitem{Chetyrkin:1981qh}
K.~G. Chetyrkin and F.~V. Tkachov, {\it {Integration by Parts: The Algorithm to
  Calculate beta Functions in 4 Loops}},
\href{http://dx.doi.org/10.1016/0550-3213(81)90199-1}{{\em Nucl. Phys.}
  {\bfseries B192} (1981) 159--204}
  [\href{http://inspirehep.net/search?p=find+Chetyrkin:1981qh}{{\ttfamily
  InSPIRE}}].
%%CITATION = NUPHA,B192,159;%%.

\bibitem{Lee:2013mka}
R.~N. Lee, {\it {LiteRed 1.4: a powerful tool for reduction of multiloop
  integrals}},
\href{http://dx.doi.org/10.1088/1742-6596/523/1/012059}{{\em J. Phys. Conf.
  Ser.} {\bfseries 523} (2014) 012059}
  [\href{http://arxiv.org/abs/1310.1145}{{\ttfamily arXiv:1310.1145}}]
  [\href{http://inspirehep.net/search?p=find+Lee:2013mka}{{\ttfamily
  InSPIRE}}].
%%CITATION = ARXIV:1310.1145;%%.

\bibitem{Peraro:2019svx}
T.~Peraro, {\it {FiniteFlow: multivariate functional reconstruction using
  finite fields and dataflow graphs}},
\href{http://dx.doi.org/10.1007/JHEP07(2019)031}{{\em JHEP} {\bfseries 07}
  (2019) 031} [\href{http://arxiv.org/abs/1905.08019}{{\ttfamily
  arXiv:1905.08019}}]
  [\href{http://inspirehep.net/search?p=find+Peraro:2019svx}{{\ttfamily
  InSPIRE}}].
%%CITATION = ARXIV:1905.08019;%%.

\bibitem{Laporta:2000dsw}
S.~Laporta, {\it {High precision calculation of multiloop Feynman integrals by
  difference equations}},
  \href{http://dx.doi.org/10.1142/S0217751X00002159}{{\em Int. J. Mod. Phys. A}
  {\bfseries 15} (2000) 5087--5159}
  [\href{http://arxiv.org/abs/hep-ph/0102033}{{\ttfamily hep-ph/0102033}}]
  [\href{http://inspirehep.net/search?p=find+Laporta:2000dsw}{{\ttfamily
  InSPIRE}}].

\bibitem{Liu:2018dmc}
X.~Liu and Y.-Q. Ma, {\it {Determining arbitrary Feynman integrals by vacuum
  integrals}},  \href{http://dx.doi.org/10.1103/PhysRevD.99.071501}{{\em Phys.
  Rev. D} {\bfseries 99} (2019) 071501}
  [\href{http://arxiv.org/abs/1801.10523}{{\ttfamily arXiv:1801.10523}}]
  [\href{http://inspirehep.net/search?p=find+Liu:2018dmc}{{\ttfamily
  InSPIRE}}].

\bibitem{Guan:2019bcx}
X.~Guan, X.~Liu, and Y.-Q. Ma, {\it {Complete reduction of two-loop
  five-light-parton scattering amplitudes}},
  \href{http://dx.doi.org/10.1088/1674-1137/44/9/093106}{{\em Chin. Phys. C}
  {\bfseries 44} (2020) 9} [\href{http://arxiv.org/abs/1912.09294}{{\ttfamily
  arXiv:1912.09294}}]
  [\href{http://inspirehep.net/search?p=find+Guan:2019bcx}{{\ttfamily
  InSPIRE}}].

\bibitem{Kotikov:1990kg}
A.~V. Kotikov, {\it {Differential equations method: New technique for massive
  Feynman diagrams calculation}},
\href{http://dx.doi.org/10.1016/0370-2693(91)90413-K}{{\em Phys. Lett.}
  {\bfseries B254} (1991) 158--164}
  [\href{http://inspirehep.net/search?p=find+Kotikov:1990kg}{{\ttfamily
  InSPIRE}}].
%%CITATION = PHLTA,B254,158;%%.

\bibitem{Caffo:2008aw}
M.~Caffo, H.~Czyz, M.~Gunia, and E.~Remiddi, {\it {BOKASUN: A Fast and precise
  numerical program to calculate the Master Integrals of the two-loop sunrise
  diagrams}},
\href{http://dx.doi.org/10.1016/j.cpc.2008.10.011}{{\em Comput. Phys. Commun.}
  {\bfseries 180} (2009) 427--430}
  [\href{http://arxiv.org/abs/0807.1959}{{\ttfamily arXiv:0807.1959}}]
  [\href{http://inspirehep.net/search?p=find+Caffo:2008aw}{{\ttfamily
  InSPIRE}}].
%%CITATION = ARXIV:0807.1959;%%.

\bibitem{Czakon:2008zk}
M.~Czakon, {\it {Tops from Light Quarks: Full Mass Dependence at Two-Loops in
  QCD}},
\href{http://dx.doi.org/10.1016/j.physletb.2008.05.028}{{\em Phys. Lett.}
  {\bfseries B664} (2008) 307--314}
  [\href{http://arxiv.org/abs/0803.1400}{{\ttfamily arXiv:0803.1400}}]
  [\href{http://inspirehep.net/search?p=find+Czakon:2008zk}{{\ttfamily
  InSPIRE}}].
%%CITATION = ARXIV:0803.1400;%%.

\bibitem{Liu:2017jxz}
X.~Liu, Y.-Q. Ma, and C.-Y. Wang, {\it {A Systematic and Efficient Method to
  Compute Multi-loop Master Integrals}},
\href{http://dx.doi.org/10.1016/j.physletb.2018.02.026}{{\em Phys. Lett.}
  {\bfseries B779} (2018) 353--357}
  [\href{http://arxiv.org/abs/1711.09572}{{\ttfamily arXiv:1711.09572}}]
  [\href{http://inspirehep.net/search?p=find+Liu:2017jxz}{{\ttfamily
  InSPIRE}}].
%%CITATION = ARXIV:1711.09572;%%.

\bibitem{Liu:2020kpc}
X.~Liu, Y.-Q. Ma, W.~Tao, and P.~Zhang, {\it {Calculation of Feynman loop
  integration and phase-space integration via auxiliary mass flow}},
  \href{http://dx.doi.org/10.1088/1674-1137/abc538}{{\em Chin. Phys. C}
  {\bfseries 45} (2021) 013115}
  [\href{http://arxiv.org/abs/2009.07987}{{\ttfamily arXiv:2009.07987}}]
  [\href{http://inspirehep.net/search?p=find+Liu:2020kpc}{{\ttfamily
  InSPIRE}}].

\bibitem{Liu:2021wks}
X.~Liu and Y.-Q. Ma, {\it {Multiloop corrections for collider processes using
  auxiliary mass flow}},
  \href{http://dx.doi.org/10.1103/PhysRevD.105.L051503}{{\em Phys. Rev. D}
  {\bfseries 105} (2022) L051503}
  [\href{http://arxiv.org/abs/2107.01864}{{\ttfamily arXiv:2107.01864}}]
  [\href{http://inspirehep.net/search?p=find+Liu:2021wks}{{\ttfamily
  InSPIRE}}].

\bibitem{Liu:2022mfb}
Z.-F. Liu and Y.-Q. Ma, {\it {Determining Feynman Integrals with Only Input
  from Linear Algebra}},
  \href{http://dx.doi.org/10.1103/PhysRevLett.129.222001}{{\em Phys. Rev.
  Lett.} {\bfseries 129} (2022) 222001}
  [\href{http://arxiv.org/abs/2201.11637}{{\ttfamily arXiv:2201.11637}}]
  [\href{http://inspirehep.net/search?p=find+Liu:2022mfb}{{\ttfamily
  InSPIRE}}].

\bibitem{Liu:2022chg}
X.~Liu and Y.-Q. Ma, {\it {AMFlow: A Mathematica package for Feynman integrals
  computation via auxiliary mass flow}},
  \href{http://dx.doi.org/10.1016/j.cpc.2022.108565}{{\em Comput. Phys.
  Commun.} {\bfseries 283} (2023) 108565}
  [\href{http://arxiv.org/abs/2201.11669}{{\ttfamily arXiv:2201.11669}}]
  [\href{http://inspirehep.net/search?p=find+Liu:2022chg}{{\ttfamily
  InSPIRE}}].

\bibitem{Jegerlehner:2000dz}
F.~Jegerlehner, {\it {Facts of life with gamma(5)}},
  \href{http://dx.doi.org/10.1007/s100520100573}{{\em Eur. Phys. J. C}
  {\bfseries 18} (2001) 673--679}
  [\href{http://arxiv.org/abs/hep-th/0005255}{{\ttfamily hep-th/0005255}}]
  [\href{http://inspirehep.net/search?p=find+Jegerlehner:2000dz}{{\ttfamily
  InSPIRE}}].

\bibitem{Kreimer:1989ke}
D.~Kreimer, {\it {The $\gamma$(5) Problem and Anomalies: A Clifford Algebra
  Approach}},  \href{http://dx.doi.org/10.1016/0370-2693(90)90461-E}{{\em Phys.
  Lett. B} {\bfseries 237} (1990) 59--62}
  [\href{http://inspirehep.net/search?p=find+Kreimer:1989ke}{{\ttfamily
  InSPIRE}}].

\bibitem{Korner:1991sx}
J.~G. Korner, D.~Kreimer, and K.~Schilcher, {\it {A Practicable gamma(5) scheme
  in dimensional regularization}},
  \href{http://dx.doi.org/10.1007/BF01559471}{{\em Z. Phys. C} {\bfseries 54}
  (1992) 503--512}
  [\href{http://inspirehep.net/search?p=find+Korner:1991sx}{{\ttfamily
  InSPIRE}}].

\bibitem{Kreimer:1993bh}
D.~Kreimer, {\it {The Role of gamma(5) in dimensional regularization}},
  [\href{http://arxiv.org/abs/hep-ph/9401354}{{\ttfamily hep-ph/9401354}}]
  [\href{http://inspirehep.net/search?p=find+Kreimer:1993bh}{{\ttfamily
  InSPIRE}}].

\bibitem{Chen:2023lus}
L.~Chen, {\it {An observation on Feynman diagrams with axial anomalous
  subgraphs in dimensional regularization with an anticommuting
  \ensuremath{\gamma}$_{5}$}},
  \href{http://dx.doi.org/10.1007/JHEP11(2023)030}{{\em JHEP} {\bfseries 2023}
  (2023) 30}
  [\href{http://arxiv.org/abs/2304.13814}{{\ttfamily arXiv:2304.13814}}]
  [\href{http://inspirehep.net/search?p=find+Chen:2023lus}{{\ttfamily
  InSPIRE}}].

\bibitem{Melnikov:2000zc}
K.~Melnikov and T.~van Ritbergen, {\it {The Three loop on-shell renormalization
  of QCD and QED}},
  \href{http://dx.doi.org/10.1016/S0550-3213(00)00526-5}{{\em Nucl. Phys. B}
  {\bfseries 591} (2000) 515--546}
  [\href{http://arxiv.org/abs/hep-ph/0005131}{{\ttfamily hep-ph/0005131}}]
  [\href{http://inspirehep.net/search?p=find+Melnikov:2000zc}{{\ttfamily
  InSPIRE}}].

\bibitem{Czakon:2007wk}
M.~Czakon, A.~Mitov, and S.~Moch, {\it {Heavy-quark production in gluon fusion
  at two loops in QCD}},
  \href{http://dx.doi.org/10.1016/j.nuclphysb.2008.02.001}{{\em Nucl. Phys. B}
  {\bfseries 798} (2008) 210--250}
  [\href{http://arxiv.org/abs/0707.4139}{{\ttfamily arXiv:0707.4139}}]
  [\href{http://inspirehep.net/search?p=find+Czakon:2007wk}{{\ttfamily
  InSPIRE}}].

\bibitem{Barnreuther:2013qvf}
P.~B\"arnreuther, M.~Czakon, and P.~Fiedler, {\it {Virtual amplitudes and
  threshold behaviour of hadronic top-quark pair-production cross sections}},
  \href{http://dx.doi.org/10.1007/JHEP02(2014)078}{{\em JHEP} {\bfseries 02}
  (2014) 078} [\href{http://arxiv.org/abs/1312.6279}{{\ttfamily
  arXiv:1312.6279}}]
  [\href{http://inspirehep.net/search?p=find+Barnreuther:2013qvf}{{\ttfamily
  InSPIRE}}].

\bibitem{Kinoshita:1962ur}
T.~Kinoshita, {\it {Mass singularities of Feynman amplitudes}},
\href{http://dx.doi.org/10.1063/1.1724268}{{\em J.Math.Phys.} {\bfseries 3}
  (1962) 650--677}
  [\href{http://inspirehep.net/search?p=find+Kinoshita:1962ur}{{\ttfamily
  InSPIRE}}].
%%CITATION = JMAPA,3,650;%%.

\bibitem{Lee:1964is}
T.~Lee and M.~Nauenberg, {\it {Degenerate Systems and Mass Singularities}},
\href{http://dx.doi.org/10.1103/PhysRev.133.B1549}{{\em Phys.Rev.} {\bfseries
  133} (1964) B1549--B1562}
  [\href{http://inspirehep.net/search?p=find+Lee:1964is}{{\ttfamily InSPIRE}}].
%%CITATION = PHRVA,133,B1549;%%.

\bibitem{ParticleDataGroup:2022pth}
{\bfseries Particle Data Group} , R.~L. Workman {\em et al.}, {\it {Review of
  Particle Physics}},  \href{http://dx.doi.org/10.1093/ptep/ptac097}{{\em PTEP}
  {\bfseries 2022} (2022) 083C01}
  [\href{http://inspirehep.net/search?p=find+ParticleDataGroup:2022pth}{{\ttfamily
  InSPIRE}}].

\bibitem{Chetyrkin:2000yt}
K.~G. Chetyrkin, J.~H. Kuhn, and M.~Steinhauser, {\it {RunDec: A Mathematica
  package for running and decoupling of the strong coupling and quark masses}},
   \href{http://dx.doi.org/10.1016/S0010-4655(00)00155-7}{{\em Comput. Phys.
  Commun.} {\bfseries 133} (2000) 43--65}
  [\href{http://arxiv.org/abs/hep-ph/0004189}{{\ttfamily hep-ph/0004189}}]
  [\href{http://inspirehep.net/search?p=find+Chetyrkin:2000yt}{{\ttfamily
  InSPIRE}}].

\bibitem{Herren:2017osy}
F.~Herren and M.~Steinhauser, {\it {Version 3 of RunDec and CRunDec}},
  \href{http://dx.doi.org/10.1016/j.cpc.2017.11.014}{{\em Comput. Phys.
  Commun.} {\bfseries 224} (2018) 333--345}
  [\href{http://arxiv.org/abs/1703.03751}{{\ttfamily arXiv:1703.03751}}]
  [\href{http://inspirehep.net/search?p=find+Herren:2017osy}{{\ttfamily
  InSPIRE}}].

\bibitem{Hoang:2013uda}
A.~H. Hoang and M.~Stahlhofen, {\it {The Top-Antitop Threshold at the ILC: NNLL
  QCD Uncertainties}},  \href{http://dx.doi.org/10.1007/JHEP05(2014)121}{{\em
  JHEP} {\bfseries 05} (2014) 121}
  [\href{http://arxiv.org/abs/1309.6323}{{\ttfamily arXiv:1309.6323}}]
  [\href{http://inspirehep.net/search?p=find+Hoang:2013uda}{{\ttfamily
  InSPIRE}}].

\bibitem{Larin:1991tj}
S.~A. Larin and J.~A.~M. Vermaseren, {\it {The alpha-s**3 corrections to the
  Bjorken sum rule for polarized electroproduction and to the Gross-Llewellyn
  Smith sum rule}},  \href{http://dx.doi.org/10.1016/0370-2693(91)90839-I}{{\em
  Phys. Lett. B} {\bfseries 259} (1991) 345--352}
  [\href{http://inspirehep.net/search?p=find+Larin:1991tj}{{\ttfamily
  InSPIRE}}].

\bibitem{Larin:1993tq}
S.~A. Larin, {\it {The Renormalization of the axial anomaly in dimensional
  regularization}},  \href{http://dx.doi.org/10.1016/0370-2693(93)90053-K}{{\em
  Phys. Lett. B} {\bfseries 303} (1993) 113--118}
  [\href{http://arxiv.org/abs/hep-ph/9302240}{{\ttfamily hep-ph/9302240}}]
  [\href{http://inspirehep.net/search?p=find+Larin:1993tq}{{\ttfamily
  InSPIRE}}].

\bibitem{Ahmed:2021spj}
T.~Ahmed, L.~Chen, and M.~Czakon, {\it {Renormalization of the flavor-singlet
  axial-vector current and its anomaly in dimensional regularization}},
  \href{http://dx.doi.org/10.1007/JHEP05(2021)087}{{\em JHEP} {\bfseries 05}
  (2021) 087} [\href{http://arxiv.org/abs/2101.09479}{{\ttfamily
  arXiv:2101.09479}}]
  [\href{http://inspirehep.net/search?p=find+Ahmed:2021spj}{{\ttfamily
  InSPIRE}}].

\bibitem{BINOSI200476}
D.~Binosi and L.~Theussl, {\it Jaxodraw: A graphical user interface for drawing
  feynman diagrams},
  \href{http://dx.doi.org/https://doi.org/10.1016/j.cpc.2004.05.001}{{\em
  Computer Physics Communications} {\bfseries 161} (2004) 76--86}
  [\href{http://inspirehep.net/search?p=find+BINOSI200476}{{\ttfamily
  InSPIRE}}].
  \url{https://www.sciencedirect.com/science/article/pii/S0010465504002115}.


\end{thebibliography}
\providecommand{\href}[2]{#2}\begingroup\raggedright\endgroup

\end{document}